# Time-dependent attractive thermal quantum force upon a Brownian free particle in the large friction regime

A. O. Bolivar*

# Time-dependent attractive thermal quantum force upon a Brownian free particle in the large friction regime

A. O. Bolivar*

Instituto Mário Schönberg de Física-Matemática-Filosofia, Ceilândia, Caixa Postal 7316, 72225-971, D.F, Brazil

We quantize the Brownian motion undergone by a free particle in the absence of inertial force (the so-called large friction regime) as described by the diffusion equation early found out by Einstein in 1905. Accordingly, we are able to come up with a time-dependent attractive quantum force $\mathcal{F}(t)$ that acts upon the Brownian free particle as a result of quantum-mechanical thermal fluctuations of a heat bath consisting of a set of quantum harmonic oscillators having the same oscillation frequency $\omega$ in thermodynamic equilibrium at temperature $T$. More specifically, at zero temperature we predict that the zero-point force is given by

$$\mathcal{F}^{(T=0)}(t) = -\frac{\omega}{(1+2\omega t)^{3/2}}\sqrt{\frac{\gamma\hbar}{2}},$$

where $\gamma$ is the friction constant with dimensions of mass per time and $\hbar$ the Planck constant divided by $2\pi$. For evolution times $t\sim 1/\omega$, $\omega\sim 10^{14}\,\mathrm{Hz}$, $\gamma\sim 10^{-10}\,\mathrm{kg/s}$, and $\hbar\sim 10^{-34}\,\mathrm{m^2 kg/s}$, we find out $\left|\mathcal{F}^{(T=0)}\right|\sim 10^{-8}\,\mathrm{N}$, which exhibits the same magnitude order as the Casimir electromagnetic quantum force, for instance. Thus, we reckon that novel quantum effects arising from our concept of time-dependent thermal quantum force $\mathcal{F}(t)$ may be borne out by some experimental set-up in nanotechnology.

PACS numbers: 05.40.-a; 05.40. Jc; 05.60. Gg

Key words: Quantum Brownian motion; strong friction limit

The centenary theory of Brownian motion has been applied to a plethora of phenomena in physics, chemistry, astronomy, electrical engineering, as well as in mathematics [1—8]. Historically, Einstein [1] came up with a mathematical description of Brownian motion in terms of the time evolution of the probability distribution function $f(x,t)$ for a free particle immersed in an environment (the so-called diffusion equation)

$$\frac{\partial f(x,t)}{\partial t} = \mathcal{D}(\infty)\frac{\partial^2 f(x,t)}{\partial x^2}, \tag{1}$$

where $x$ is the position of the Brownian particle as inertial forces upon it may be negligible (the large friction domain). The diffusion constant $\mathcal{D}(\infty)$, calculated at the long time regime, with dimensions of $[length^2 \times time^{-1}]$ may be written down as

$$\mathcal{D}(\infty) = \frac{\mathcal{E}(\infty)}{\gamma}, \tag{2}$$

where the steady quantity $\mathcal{E}(\infty)$ having dimensions of energy is termed the diffusion energy responsible for the Brownian motion, whereas the damping constant $\gamma$ exhibiting dimensions of mass per time is related to mechanical properties inherent in the interaction between the Brownian massive particle and the environmental particles.

On the condition that the diffusion energy $\mathcal{E}(\infty)$, which comes from the Brownian particle, can be identified with the thermal energy $k_B T$ coming from the heat bath in thermodynamic equilibrium at temperature $T$, i.e.,

$$\mathcal{E}(\infty) = k_B T, \tag{3}$$

we readily obtain the Einstein's fluctuation-dissipation relation [1]

$$\mathcal{D}(\infty) = \frac{k_B T}{\gamma}. \tag{4}$$

The thermodynamic constant $k_B$ is dubbed the Boltzmann constant. Assumption (3) is a form of expressing the principle of energy conservation.

On the basis of the diffusion equation (1) and the diffusion constant (4) Einstein [1] derived the mean square displacement as

$$\Delta X(t) = \sqrt{2\frac{k_B T}{\gamma}t}, \tag{5}$$

which is the physically measurable quantity in the theory of Brownian motion. For $t = 1\text{s}$, $k_B \sim 10^{-23}\text{m}^2\text{kgs}^{-2}\text{K}^{-1}$, $T \sim 300\text{K}$, and $\gamma \sim 10^{-9}\text{kgs}^{-1}$, Einstein obtained a value of the order of $\Delta X \sim \mu\text{m}$.

It is worth noticing that no diffusive motion, $\mathcal{D}(\infty) = 0$, and no fluctuation, $\Delta X(t) = 0$, are predicted at zero temperature. Yet, as far as quantum effects are concerned, both the Brownian dynamics (1) and the heat bath's thermal energy (3) must depend on the quantum action ħ, the Planck's constant divided by $2\pi$. Therefore, the diffusion energy (3) and in consequence both the constant diffusion (4) and the mean square displacement (5) cannot vanish at the zero-temperature realm on account of the existence of the zero-point energy.

Although the environment can be quantum-statistically treated as a heat bath consisting of a set of harmonic oscillators in thermal equilibrium at temperature $T$ [9], the quantization process of the Brownian dynamics (1) holds still a *terra incognita*. We reckon that inquiring into this outstanding physical problem can shed some light on the quantum nature of Brownian motion thereby implying ongoing advances in nanotechnology, for example.

The aim of the present paper is then to take up the physical-mathematical problem of quantizing the diffusion equation (1) reckoning with the quantum nature of the heat bath. To this end, we start from it, with (2), at points $x_1$ and $x_2$, i.e.,

$$\frac{\partial \chi(x_1,t)}{\partial t} = \frac{\overline{\mathcal{E}}(\infty)}{\gamma}\frac{\partial^2 \chi(x_1,t)}{\partial x_1^2} \tag{6}$$

and

$$\frac{\partial \chi(x_2,t)}{\partial t} = \frac{\overline{\mathcal{E}}(\infty)}{\gamma}\frac{\partial^2 \chi(x_2,t)}{\partial x_2^2}\,. \tag{7}$$

Solutions $\chi(x_1,t)$ and $\chi(x_2,t)$ are deemed to be associated with the diffusion energy $\overline{\mathcal{E}}(\infty) = 2\mathcal{E}(\infty)$. By multiplying (6) by $\chi(x_2,t)$ and (7) by $\chi(x_1,t)$ and adding the resulting equations, we obtain

$$\frac{\partial \xi(x_1,x_2,t)}{\partial t} = \frac{2\mathcal{E}(\infty)}{\gamma}\left[\frac{\partial^2 \xi(x_1,x_2,t)}{\partial x_1^2} + \frac{\partial^2 \xi(x_1,x_2,t)}{\partial x_2^2}\right], \tag{8}$$

where $\xi(x_1,x_2,t) = \chi(x_1,t)\chi(x_2,t) = \sqrt{f(x_1,t)f(x_2,t)}$.

By performing the following change of variables into configuration space, $(x_1,x_2) \longmapsto (x,\eta\hbar)$,

$$x_1 = x - \frac{\eta\hbar}{2}, \tag{9}$$

$$x_2 = x + \frac{\eta\hbar}{2}, \tag{10}$$

the classical equation (8) changes into the quantum equation of motion

$$\frac{\partial\rho(x,\eta,t)}{\partial t}=\frac{\mathcal{E}_{\hbar}(\infty)}{\gamma}\left[\frac{\partial^2\rho(x,\eta,t)}{\partial x^2}+\frac{4}{\hbar^2}\frac{\partial^2\rho(x,\eta,t)}{\partial\eta^2}\right], \tag{11}$$

where we have replaced the solution $\xi(x_1,x_2,t)$ with $\rho(x,\eta,t)$ and employed the subscript notation in $\mathcal{E}_{\hbar}(\infty)$ to disclose the quantum nature of the diffusion energy which turns out to be now expressed in terms of the Planck constant ħ. The variable $\eta$ displays dimensions of inverse of linear momentum.

The geometric meaning of the quantization conditions (9) and (10) is related to the existence of a minimal distance between the points $x_1$ and $x_2$ due to the quantum nature of space, $x_2-x_1=\eta\hbar$, such that in the classical limit $\hbar\to 0$, physically interpreted as $|\eta\hbar|\ll|x_2-x_1|$, the result $x_2=x_1=x$ can be readily recovered.

By making use of the Fourier transform

$$W(x,p,t)=\frac{1}{2\pi}\int_{-\infty}^{\infty}\rho(x,\eta,t)e^{ip\eta}\,d\eta, \tag{12}$$

which changes the variables from quantum configuration space $(x,\eta\hbar)$ onto quantum phase space $(x,p;\hbar)$, the quantum dynamics (11) turns out to be written down as

$$\frac{\partial W(x,p,t)}{\partial t}=\frac{\mathcal{E}_{\hbar}(\infty)}{\gamma}\frac{\partial^2 W(x,p,t)}{\partial x^2}-\frac{4\mathcal{E}_{\hbar}(\infty)}{\hbar^2\gamma}p^2W(x,p,t). \tag{13}$$

Because the exponential factor $e^{ip\eta}$ in (12) is to be a dimensionless term, it follows that $p$ is to have dimensions of linear momentum. Thus both quantum equations of motion (11) and (13) arise from our method of quantizing the Brownian dynamics (1).

We now assume that the environment is made up of a set of harmonic oscillators having the same oscillation frequency $\omega$, so that the mean energy of this quantum heat bath after attaining the thermodynamic equilibrium at temperature $T$ is given by the Boltzmann—Maxwell statistics [9]

$$\overline{E}=\frac{\omega\hbar}{2}\left(\frac{e^{\frac{\omega\hbar}{k_BT}}+1}{e^{\frac{\omega\hbar}{k_BT}}-1}\right)=\frac{\omega\hbar}{2}\coth\left(\frac{\omega\hbar}{2k_BT}\right), \tag{14}$$

where the ħ-dependent energy, $\omega\hbar/2$, corresponds to the zero-point energy inherent in the quantum heat bath at zero temperature and $k_BT$ the classical thermal energy of the quantum heat bath at high temperatures $T\gg\omega\hbar/2k_B$.

On the condition that the Brownian particle's quantum diffusion energy $\mathcal{E}_{\hbar}(\infty)$ can be identified with the heat bath's quantum thermal energy $\overline{E}$, we obtain

$$\mathcal{E}_{\hbar}(\infty) = \frac{\omega\hbar}{2}\coth\left(\frac{\omega\hbar}{2k_B T}\right), \tag{15}$$

from which the quantum diffusion coefficient can be derived as

$$\mathcal{D}_{\hbar}(\infty) = \frac{\mathcal{E}_{\hbar}(\infty)}{\gamma} = \frac{\omega\hbar}{2\gamma}\coth\left(\frac{\omega\hbar}{2k_B T}\right). \tag{16}$$

Our quantum phase-space diffusion equation (13) may be solved starting from the non-thermal initial condition

$$W(x,p,t=0) = \frac{1}{\pi\hbar} e^{-\left(\frac{ap^2}{\hbar}+\frac{x^2}{\hbar a}\right)} \tag{17}$$

leading to the distribution $F(x,p) = \delta(x)\delta(p)$ in the classical limit, $\hbar \to 0$. The constant $a$ has dimensions of time per mass, hence we may set $a = 1/\gamma$. The quantum steady probability distribution function (17) generates the fluctuations $\Delta P(0) = \sqrt{\gamma\hbar/2}$ and $\Delta X(0) = \sqrt{\hbar/2\gamma}$, which comply with the Heisenberg indeterminacy principle $\Delta P(0)\,\Delta X(0) = \hbar/2$.

The time-dependent solution to (13), with (15), reads

$$W(x,p,t) = \frac{1}{\pi\hbar} e^{-\frac{4B_{\hbar}(t)}{\hbar^2}p^2} e^{\frac{-x^2}{4B_{\hbar}(t)}}, \tag{18}$$

where

$$B_{\hbar}(t) = \frac{\hbar}{4\gamma} + \frac{\omega\hbar t}{2\gamma}\coth\left(\frac{\omega\hbar}{2k_B T}\right).$$

Solution (18) leads then to both the quantum mean square displacement

$$\Delta X(t) = \sqrt{\frac{\hbar}{2\gamma}\left[1 + 2t\omega\coth\left(\frac{\omega\hbar}{2k_B T}\right)\right]} \tag{19}$$

and the quantum mean square momentum

$$\Delta P(t) = \sqrt{\frac{\gamma\hbar}{2}}\frac{1}{\sqrt{1 + 2t\omega\coth\left(\frac{\omega\hbar}{2k_B T}\right)}}, \tag{20}$$

which in turn do fulfill the Heisenberg constraint: $\Delta P(t)\Delta X(t) = \hbar/2$.

The quantum mean square displacement (19) gives rise to the quantum velocity $\mathcal{V}(t) = d\Delta X(t)/dt$, while the quantum mean square momentum (20) generates the *time-dependent attractive thermal quantum force* $\mathcal{F}(t) = d\Delta P(t)/dt$ given by

$$\mathcal{F}(t) = -\sqrt{\frac{\gamma\hbar}{2}}\frac{\omega\coth\left(\frac{\omega\hbar}{2k_BT}\right)}{\left[1+2t\omega\coth\left(\frac{\omega\hbar}{2k_BT}\right)\right]^{3/2}}, \tag{21}$$

ranging from $\mathcal{F}(t=0) = -\sqrt{\gamma\hbar/2}\,\omega\coth(\omega\hbar/2k_BT)$ to $\mathcal{F}(\infty) = 0$.

From the Heisenberg relation $\Delta P(t)\Delta X(t) = \hbar/2$ it is readily to show that the quantities $\mathcal{F}(t)$ and $\mathcal{V}(t)$ are connected via the relationship $\mathcal{F}(t) = -[\Delta P(t)/\Delta X(t)]\mathcal{V}(t)$. From the physical viewpoint the attractive quantum force (21) is exerted upon the Brownian free particle as a result of quantum-mechanical thermal fluctuations of the heat bath as well as the quantum Brownian dynamics (13).

The thermal quantum force (21) at zero temperature reads

$$\mathcal{F}^{(T=0)}(t) = -\sqrt{\frac{\gamma\hbar}{2}}\frac{\omega}{(1+2\omega t)^{3/2}}. \tag{22}$$

For evolution times $t\sim1/\omega$, $\omega\sim10^{14}$ Hz, $\gamma\sim10^{-10}$kg/s, and $\hbar\sim10^{-34}\mathrm{m}^2$kg/s, we find out the following magnitude: $|\mathcal{F}^{(T=0)}|\sim10^{-8}$N. Under these conditions the mean square displacement (19) is of the order of $\Delta X^{(T=0)}\sim10^{-12}$m. Thereby, it is worth underscoring that our quantum force (21) may exhibit the same magnitude order as the Casimir force [10,11], for instance. Consequently, it may be borne out experimentally.

The high temperature regime should be physically interpreted as the temperature $T$ is deemed to be too large in comparison with the quantum temperature $T_\mathrm{q} = \omega\hbar/2k_B$ , i.e., $T \gg \omega\hbar/2k_B$, such that $\coth(\omega\hbar/2k_BT)\sim2k_BT/\omega\hbar$. Thus, the mean square displacement (19) becomes

$$\Delta X(t) = \sqrt{\frac{\hbar}{2\gamma}+\frac{2k_BTt}{\gamma}}, \tag{23}$$

where the $\hbar$-dependence comes from the quantum dynamics (13) starting from the initial condition (17) while the $\hbar$-independent term is due to the classical nature of the heat bath. On the other hand, our thermal quantum force (21) at high temperature turns out to be given by

$$\mathcal{F}(t) = -\frac{\hbar k_BT\sqrt{2\gamma}}{(\hbar+4tk_BT)^{3/2}}. \tag{24}$$

For $\gamma\sim10^{-10}$kg/s at $T\sim10^3$K, and $t=0$, Eq. (23) and Eq. (24) lead to $\Delta X(0)\sim10^{-12}$m and $|\mathcal{F}(t=0)|\sim10^{-8}$N, respectively.

The classical limit of the Brownian dynamics should be interpreted as the evolution time scale $t$ is considered too large in comparison with the quantum time scale $t_{\mathrm{q}} = \omega^{-1} = \hbar/4k_B T$, i.e., $t \gg \hbar/4k_B T$. At 100 K, $t_{\mathrm{q}} \sim 10^{-13}$s, for instance. So Eq. (23) reduces to the Einstein's classical mean square displacement (5). Taking into account the Einstein's particle [1] characterized by $\gamma \sim 10^{-10}$kg/s at $t = 1$s, Eq.(24) leads to the magnitude $|\mathcal{F}(t = 1\mathrm{s})| \sim 10^{-30}$N. In other words, our quantum force (21) virtually vanishes in the classical realm characterized by $T \gg T_{\mathrm{q}}$ and $t \gg t_{\mathrm{q}}$.

In summary, in this paper we have examined the quantum Brownian motion of a free particle in the absence of inertial force (the so-called strong friction case). Our crucial finding is the concept of time-dependent attractive thermal quantum force (21) exerted upon the quantum Brownian free particle by the quantum-mechanical heat bath. We reckon that this intrinsically quantum physical effect can be experimentally confirmed since at zero temperature it may have the same magnitude order as the Casimir force, for instance. Moreover, our upshot (21) may play a pivotal role in quantum nanothermomechanical systems at the low-temperature range $T \sim \omega\hbar/2k_B$, as it virtually dies out at high temperatures.

I thank Professor Maria Carolina Nemes for the scientific support and FAPEMIG (Fundação de Amparo à Pesquisa do Estado de Minas Gerais) for the financial support.

*Present address: Departamento de Física, Universidade Federal de Minas Gerais, Caixa Postal 702, 30123-970, Belo Horizonte, Minas Gerais, Brazil. Electronic mail: bolivar@cbpf.br.